\documentclass[aps,prb,twocolumn,floatfix]{revtex4-2}

\usepackage{amssymb}
\usepackage{amsmath}
\usepackage{graphicx}

\begin{document}

\title{Determination of the London penetration depth with the tunnel diode oscillator technique}

\author{G.~P.~Mikitik}
\affiliation{B.~Verkin Institute for Low Temperature Physics \&
Engineering of the National Academy of Sciences of Ukraine, Kharkiv 61103, Ukraine}

\begin{abstract} Using a distribution of the Meissner currents over the surface of an infinitely long superconducting slab with a rectangular cross section, the magnetic moment of the slab is calculated, taking into account corrections associated with a small but finite value of the London penetration depth $\lambda$. Since these corrections determine the shift of the resonant frequency in the tunnel diode oscillator technique, formulas for determination of $\lambda$ within this technique are derived for the slab. These formulas are valid for any aspect ratio of its cross section, and they differ from those that are often used in analyzing experimental data. Namely, it is shown that sharp edges of the slab can cause the large frequency shift proportional to the change in the value of $\lambda^{2/3}$. Although this result complicates the extraction of a temperature dependence of $\lambda$ from the frequency shift, it also opens up new possibilities in determining the London penetration depth. In particular, under certain conditions, it is possible not only to measure the changes in $\lambda$ with the temperature, but also to estimate its absolute value.
\end{abstract}

\maketitle

\section{Introduction}

A dependence of the  London penetration depth $\lambda$ on the temperature $T$ can give an important information  on the pairing state of electrons in superconductors \cite{proz06}. For example, such information were obtained for YBa$_2$Cu$_3$O$_{6.95}$  \cite{hardy,carrington},  Fe-based \cite{proz11,putzke14,shen}, filled-skutterudite \cite{juraszek20}, and  heavy-fermion
\cite{yamashita,takenaka,pang,ishihara} superconductors. Below we will discuss the tunnel diode oscillator technique \cite{carrington,proz00,giannetta} that is widely used (see, e.g., Refs.~\cite{yamashita,takenaka,pang,ishihara,
klein,smylie,prib,shang18,kim1,radmanesh,kim,ishihara1,wang21,su,nie,duan,chapai})  to measure the changes in $\lambda$, $\Delta\lambda\equiv \lambda(T)-\lambda(T_{\rm min})$, when the temperature $T$ of a superconductor increases from some starting value $T_{\rm min}$.
In this technique, the changes in the London penetration depth manifest themselves in the shift $\Delta f\equiv f(T)-f(T_{\rm min})$ of the resonant frequency $f$ of a tank circuit, and $\Delta \lambda$ can be found from the relation \cite{proz00}:
\begin{eqnarray}\label{1}
\frac{\Delta f}{\delta f}=-\frac{\Delta\lambda}{R},
\end{eqnarray}
where  $\delta f$ is the change of the frequency when the sample is  removed from the coil of the circuit (this $\delta f$ is measured directly), and the length $R$ determines the dependence of the magnetic moment $M$ of the superconducting sample in the Meissner state on $\lambda$ at small values of the London penetration depth,
 \[
 M(\lambda)\approx M(0)\!\left(1-\frac{\lambda}{R}\right),
 \]
where $M(0)$ is the magnetic moment in the limit $\lambda\to 0$.
For an infinite superconducting  plate in the parallel magnetic field, $R$ is a half of the thickness of the plate, whereas in general case, $R$ depends on the distribution of the Meissner currents over the surface of the superconductor and therefore, on its shape. In most experiments, the sample is a parallelepiped, and  approximate formulas for $R$ were proposed for this practically important situation \cite{proz00,proz_pr-appl21}.

On the other hand, for an infinitely long slab with a rectangular cross section, a distribution of the Meissner currents over its surface is known in the limit $\lambda\to 0$  \cite{Meissner}. In this paper, using this strict result, a correction to $M(0)$ is calculated for small but finite $\lambda$, i.e., when this $\lambda$ is much less than the width $2w$ and the thickness $d$ of the slab. This calculation gives the value of $R$ for an arbitrary aspect ratio $d/2w$ of the sample. Moreover, it is shown in this paper that the edges of the slab give a qualitatively different correction to $M(\lambda)$. This correction is proportional to $\lambda^{2/3}$, and it exceeds the linear-in-$\lambda$ contribution to $M(\lambda)$ for thin samples with $d\ll 2w$.

 \begin{figure}[t] % %%%%%%%%%%%%%%%%%%%%%%%%%%%%%%%%%%%%%
 \centering  \vspace{+9 pt}
\includegraphics[trim=16mm 55mm 10mm 0mm,scale=.37]{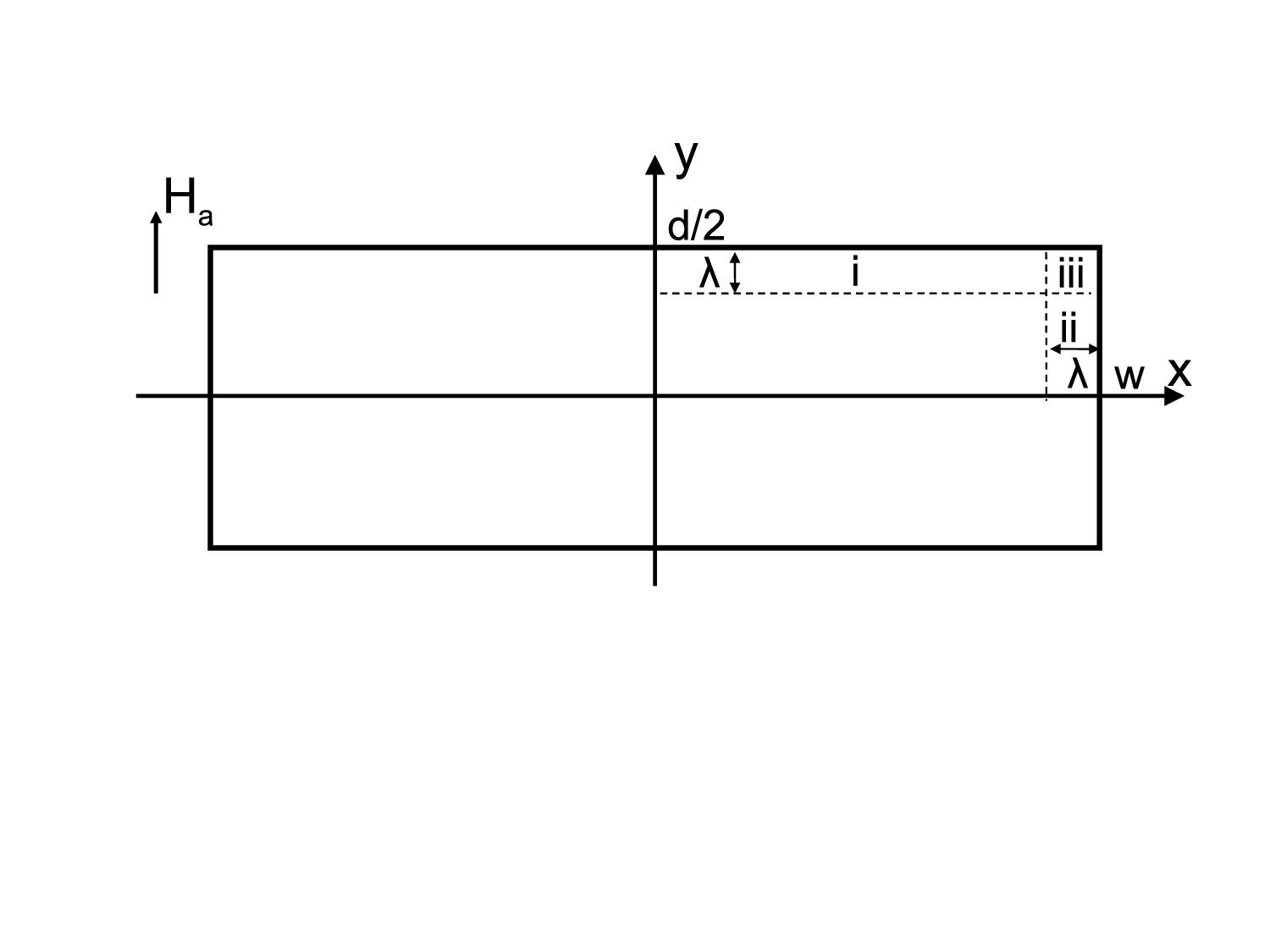}  %scale=39
\caption{\label{fig1} The rectangular cross section of the infinitely long superconducting slab of the width $2w$ and of the thickness $d$. The magnetic field $H_a$ is applied along the $y$ axis. The dashed lines schematically indicate the three parts of the surface layer in the upper right quarter of the slab ($x>0$, $y>0$). In this layer, the Meissner currents flow along the $z$ axis.
 } \end{figure}   %%%%%%%%%%%%%%%%%%%%%%%%%%%%%%%%%%%%%%%%%%

In Sec.~\ref{II} we present the distribution of the Meissner currents in the infinitely long slab with the rectangular cross section.
Using this distribution, in Sec.~\ref{III}, the magnetic moment of the slab is calculated, taking into account corrections determined by $\lambda$. The obtained results are compared with approximate formulas of Prozorov et al. \cite{proz00,proz_pr-appl21}.
In Sec.~\ref{IV}, the above-mentioned frequency shift caused by a change in $\lambda$ is analyzed, and additional possibilities for determination of $\lambda$ are discussed. The obtained results are briefly summarized in Sec.~\ref{V}. Appendices  contain some mathematical details. In particular, the case of the elliptic cylinder in the transverse magnetic field is discussed in Appendix \ref{B}.

\section{Slab in the Meissner state} \label{II}

Consider a superconducting slab of a rectangular cross section of width $2w$ ($-w\le x\le w$) and thickness $d$ ($-d/2 \le y \le d/2$), Fig.~\ref{fig1}. The slab infinitely extends in the $z$ direction. The applied magnetic field ${\bf H}_a=(0,H_a,0)$ is directed along the thickness of the slab. Throughout this article it is assumed that $d, w \gg \lambda$.

For a superconductor in the Meissner state, the magnetic field ${\bf H}$ near the sample is tangential to its surface in the limit $\lambda\to 0$. This ${\bf H}(x,y)$ outside an infinitely long superconductor can be found by a conformal mapping, whereas the Meissner sheet currents $J_M=J_z$ flowing  on the surface can be derived from the relation,
${\bf J}_M=[{\bf n}\times {\bf H}]$, where ${\bf n}$ is the outward normal to the surface of the sample  at the point of interest \cite{LL}.
For the slab, the Meissner currents were found many years ago \cite{Meissner} (the appropriate mapping was detailed in the Supplemental Material to paper \cite{prb21}). Below we present the results of Refs.~\cite{Meissner,prb21} that are used in the subsequent sections.

Taking into account the symmetry of the slab, we consider the quarter of its surface composed of the two segments $0\le x \le w$, $y=d/2$ and $x=w$, $0\le y\le d/2$. It is convenient to parametrize this quarter by the single variable $u$ changing from $0$ to $1/\sqrt{1- m}$ \cite{prb21} where $m$ is a constant parameter, $0\le m \le 1$. The value of $m$ is determined by the aspect ratio of the slab, $d/2w$,
\begin{eqnarray} \label{2}
\frac{d}{2w}=\frac{f(1,m)}{f(1,1-m)},
\end{eqnarray}
where the function $f(u,m)$,
\begin{eqnarray} \label{3}
f(u,m)\equiv m\int_0^{u}\!\!\frac{\sqrt{1-v^2}}{\sqrt{1-mv^2}}\,dv,
\end{eqnarray}
can be expressed in terms of the elliptic integrals \cite{Meissner,prb21}. At $d\ll w$,  relation (\ref{2}) gives
\begin{eqnarray} \label{4}
m\approx \frac{2d}{\pi w}.
\end{eqnarray}
The upper surface of the slab ($0\le x\le w$, $y=d/2$) and the upper part of the lateral surface, ($x=w$, $0\le y\le d/2$)  are  described as follows:
\begin{eqnarray} \label{5}
\frac{x}{w}&=&\frac{f(u,1-m)}{f(1,1-m)}, \\
\frac{2y}{d}&=&\frac{f(s(u),m)}{f(1,m)}, \label{6}
\end{eqnarray}
where
\begin{eqnarray*}
s(u)=\sqrt{\frac{1-(1-m)u^2}{m}}.
\end{eqnarray*}
In Eq.~(\ref{5}), the parameter $u$ lies in the interval $0\le u \le 1$, and the points $u= 1$ correspond to the upper right corner of the slab, $(w,d/2)$. On the other hand, $1\le u \le 1/\sqrt{1-m}$ in  Eq.~(\ref{6}), and $u=1/\sqrt{1-m}$ at the
equatorial point of the slab, $(w,0)$.

The Meissner sheet currents in the whole interval $0\le u \le 1/\sqrt{1-m}$ (i.e., on the upper and lateral
surfaces of the slab) are described by one and the same formula:
\begin{eqnarray} \label{7}
J_M(u)=\frac{uH_a}{\sqrt{|1-u^2|}}.
\end{eqnarray}

Formulas (\ref{2})--(\ref{7}) enable one to calculate the current $I_{\rm ltr}$ flowing on each of the lateral surfaces of the slab \cite{prb21},
\begin{eqnarray}  \label{8}
I_{\rm ltr}= \frac{dH_a\sqrt{m}}{f(1,m)}=\frac{2wH_a\sqrt{m}}{f(1,1-m)},
\end{eqnarray}
and the magnetic moment $M_y$ of the slab (per its unit length) \cite{Meissner},
\begin{eqnarray}  \label{9}
M_y=-\frac{\pi H_aw^2(1-m)}{[f(1,1-m)]^2}=-\frac{\pi H_adw(1-m)}{2f(1,1-m)f(1,m)}.~~
\end{eqnarray}
As to the current $I_{\rm up}$ flowing on the upper plane of the slab in the interval $0\le x\le w$, formulas (\ref{3}), (\ref{5}), and (\ref{7}) give
\begin{eqnarray}  \label{10}
I_{\rm up}\!=\!\frac{wH_a\arcsin(\sqrt{1-m})}{f(1,1-m)}=\! \frac{dH_a\arcsin(\sqrt{1-m})}{2f(1,m)}.~~~~
\end{eqnarray}
The two representations of each of formulas (\ref{8})-(\ref{10}) follow from relation (\ref{2}).

In the limit $|1-u^2|\ll m$ (i.e., when $l\equiv w-x\ll d, w$ or  $l\equiv (d/2)-y\ll d, w$), the surface current diverges like $l^{-1/3}$ near the corners of the slab \cite{Meissner,prb21,jetp13}. In this limiting case, formulas (\ref{2})-(\ref{7}) lead to the expression \cite{prb21,m-sh24},
\begin{eqnarray} \label{11}
J_M\!\approx\! H_a\!\left(\!\frac{(1-m)d}{6\sqrt{m}f(1,m)\,l}\! \right)^{\!1/3},~~
\end{eqnarray}
which is valid for the slab of an arbitrary aspect ratio.  For the thin strips, expression (\ref{11}) is further simplified since  $f(1,m)\approx \pi m/4$ at $m\ll 1$. The divergence of the current in Eq.~(\ref{11}) should be cut off at $l\lesssim \lambda$, and the current density throughout the corner region ($w-\lambda \le x \le w$, $(d/2)-\lambda \le y\le d/2$) is approximately constant, $j_{\rm crn}(x,y)\sim J_M(x=w-\lambda)/\lambda$,
\begin{eqnarray} \label{12}
j_{\rm crn}\!\sim\!\frac{H_a}{\lambda }\!\left(\!\frac{(1-m)d}{6\sqrt{m}f(1,m)\,\lambda}\! \right)^{\!1/3}.~~
\end{eqnarray}
This prescript for the cut-off of $J_M$ is explained as follows: The conformal mapping was obtained under the assumption that the magnetic field does not penetrate into the sample ($\lambda\to 0$). The penetration of the magnetic field at the small depth $\lambda$ does not change the shape of the flat parts of the surface of the slab and is unimportant for this mapping. However, the penetration near the corners of its cross section leads to the rounding of the corners at the distance $\lambda$ from their vertices. This rounding suppresses the divergence of the Meissner  currents.

\section{Magnetic moment of the slab}\label{III}

As it was mentioned in Introduction, determination of the London penetration depth in the tunnel diode oscillator technique is based on a dependence of the magnetic moment of the superconductor in the Meissner state on $\lambda$. However, formula (\ref{9}) for the magnetic moment $M_y$ was derived in the limit $\lambda\to 0$. We will now find the corrections associated with small $\lambda$ to this result, neglecting all terms proportional to $\lambda^n$ if $n>1$.

\subsection{Magnetic moment for small $\lambda$}\label{IIIA}

For a nonzero $\lambda$, the magnetic field inside a superconductor is described by the London equation, whereas it meets the Laplace equation outside the sample. At the boundary of the superconductor  the field has to be continuous. At small $\lambda$, the distribution of the magnetic fields and currents in the superconductor can be found with the approach which is usually used in the calculation of the high-frequency magnetic field penetrating into a conductor (see \S 45 in Ref.~\cite{LL}). In other words, due to the smallness of $\lambda$, the magnetic field inside the superconductor can be calculated using the London equation for a semi-infinite medium bounded by a plane, outside which the field has a given constant value. This value for the slab coincides with a local surface magnetic field $H_t=J_M(x,y)$ described by formulas of Sec.~\ref{II}. Then, it follows from the London equation that the current densities $j_z$ near a point $(x,d/2)$ on the upper surface of the slab and  near a  point $(w,y)$ on its lateral surface are described by the following expressions:
\begin{eqnarray}\label{13}
j_z(x,y)&=&\frac{J_M(x,d/2)}{\lambda} \exp\left(-\frac{0.5d-y}{\lambda}\right),  \\  j_z(x,y)&=&\frac{J_M(w,y)}{\lambda}\exp\left(-\frac{w-x}{\lambda}\right), \nonumber
\end{eqnarray}
respectively. Knowing $J_M$ on the surface of the superconductor   and hence, the current density $j_z(x,y)$, the total magnetic moment $m_y$ of the slab can be calculated with the formula:
\begin{eqnarray}\label{14}
 m_y=\frac{1}{2}\int (zj_x-xj_z) dxdydz
 =-\int xj_z dxdydz,~~
 \end{eqnarray}
where we have taken into account that the currents $j_x$ flowing near the far edges of the slab (i.e., at large $|z|$) give the same contribution to $m_y$ as the currents $j_z$ \cite{br94,br04} (see also Appendix \ref{A}).

Consider  the Meissner currents $j_z$ flowing near the surface of the quarter of the slab ($x>0$, $y>0$) in the layer of the thickness  $\sim \lambda$. Let us divide this layer into the three parts (Fig.~\ref{fig1}): (i) the part adjoining the upper surface at $0\le x\le (w-\lambda)$ (ii) the part near the  right lateral surface at $0\le y\le (0.5d-\lambda)$, and (iii) the corner region: $(w-\lambda) \le x\le w$, $(0.5d-\lambda)\le y\le d/2$. The magnetic moment (per a unit length of the slab) generated by $j_z$ in the part (i) can be written in the form,
 \[
 M_{\rm up}(\lambda)=M_{\rm up}(0)+\frac{w}{2}I_{\rm crn},
 \]
where $M_{\rm up}(0)=-(1/2)\int_0^wxJ_M(x,d/2)dx$ is this part of the moment at $\lambda=0$, and the second term takes into account that the region (i) does not reach the point $x=w$, i.e., $I_{\rm crn}=\int_{w-\lambda}^wJ_M(x,d/2)dx$ is the Meissner current  integrated over the small interval $w-\lambda\le x\le w$. The direct calculation of $I_{\rm crn}$ with Eq.~(\ref{7}) gives $I_{\rm crn}=(3/2)j_{\rm crn}\lambda^2$ where $j_{\rm crn}$ is defined by Eq.~(\ref{12}). Note that $M_{\rm up}$ depends on $\lambda$ only via $I_{\rm crn}\propto \lambda^{2/3}$, whereas distribution (\ref{13}) has no effect on $M_{\rm up}(\lambda)$. Another situation occurs for the part (ii) of the layer. Using the above distribution of $j_z(x,y)$ near the lateral surface, we obtain the following contribution of $j_z$, flowing in the part (ii), to the magnetic moment:
  \[
  M_{\rm ltr}(\lambda)\!=\!-\frac{1}{2}(w-\lambda)(\frac{I_{\rm ltr}}{2}-I_{\rm crn})\!\approx\!M_{\rm ltr}(0)+\frac{wI_{\rm crn}}{2}+\frac{\lambda I_{\rm ltr}}{4},
  \]
where
 \[M_{\rm ltr}(0)=-(w/2)\int_0^{d/2}J_M(w,y)dy=-wI_{\rm ltr}/4 \]
is this part of the moment at $\lambda=0$, and $I_{\rm ltr}$ is described by Eq.~(\ref{8}). Note that the region (ii) produces not only the correction $wI_{\rm crn}/2$, but also the term $\lambda I_{\rm ltr}/4$ that is linear in $\lambda$. As to the corner region (iii), it gives,
  \[
  M_{\rm crn}=-\frac{w}{2}j_{\rm crn}\lambda^2=-\frac{1}{3}wI_{\rm crn}. \]
Then, the total contribution of the currents $j_z$ to the magnetic moment $m_y$ of the slab is equal to $4(M_{\rm up}+M_{\rm ltr}+M_{\rm crn})(L-2\lambda)$, and according to Eq.~(\ref{14}), this contribution must be doubled to obtain  $m_y$,
\begin{eqnarray}\label{15}
 m_y(\!\!&\lambda&\!\!)=2(L-2\lambda)\!\left[\frac{M_y(0)}{2}+\lambda I_{\rm ltr} + \frac{8}{3}wI_{\rm crn}\right] \\
 &=&-\frac{\pi (1-m)H_aV}{4f(1,1-m)f(1,m)}\!\left[1-\frac{2\lambda}{L} -\frac{\lambda}{R}-\frac{\lambda^{2/3}}{R_{\rm crn}^{2/3}}\right], \nonumber
 \end{eqnarray}
where $M_y(0)/2=4(M_{\rm up}(0)+M_{\rm ltr}(0))$, $M_y(0)$ is given by formula (\ref{9}), $L$ is the length of the slab, $V=2wdL$ is its volume, $(\lambda/R)\equiv2\lambda I_{\rm ltr}/|M_y(0)|$, and  $(\lambda/R_{\rm crn})^{2/3}\equiv 16wI_{\rm crn}/(3|M_y(0)|)$. Here we have defined the two effective sizes of the slab, $R$ and $R_{\rm crn}$. The first size is
\begin{eqnarray}\label{16}
 R=\frac{|M_y(0)|}{2I_{\rm ltr}}=w\frac{\pi(1-m)}{4f(1,1-m)\sqrt{m}},
 \end{eqnarray}
whereas the second size $R_{\rm crn}$ is associated with the edges of the slab,
  \begin{eqnarray}\label{17}
  R_{\rm crn}\!=\!\left(\frac{3|M_y(0)|\lambda^{2/3}}{16wI_{\rm crn}}\right)^{3/2}\!\!\!\!\!\!=\! w\frac{\sqrt{6}\pi^{3/2}m^{1/4}(1-m)}{32[f(1,1-m)]^{5/2}}.~~~
  \end{eqnarray}
It is also convenient to introduce the effective demagnetization factor $N$, which determines the linear dependence of $m_y(0)$ on $H_a$ in  formula (\ref{15}) at $\lambda=0$,
 \begin{eqnarray}\label{18}
 m_y(0)&=& -\frac{H_aV}{1-N}, \nonumber \\
\frac{1}{1-N}&=&\frac{\pi(1-m)}{4f(1,1-m)f(1,m)}.
  \end{eqnarray}
Note that in Eq.~(\ref{15}), we have multiplied the magnetic moment calculated per the unit length of the slab by $L-2\lambda$, which is the length of the region where the currents $j_z$ flow. However, near the edges of the slab [at the distance $\sim{\rm max}(d,2w)$ from these edges], formulas (\ref{7})-(\ref{9}) become inaccurate. Hence, the relative accuracy of all the terms in formula (\ref{15}) is of the order of ${\rm max}(d,2w)/L$.

The currents flowing near the smooth surface of an arbitrary-shaped  superconductor in the layer of the thickness $\lambda$ generally renormalize the surface magnetic field as compared to its value  in the limit $\lambda\to 0$. The renormalized field $\tilde H_t(0)$ at the surface is  described by the formula,
 \begin{eqnarray}\label{19}
 \tilde H_t(0)=H_t(0)+\lambda\frac{\partial H_t(t)}{\partial t}|_{t=0},
   \end{eqnarray}
where $t$ is a local coordinate perpendicular to the surface at a given point ($t=0$ at the surface), and $H_t(t)$ is tangential magnetic field obtained outside the superconductor in the limit $\lambda\to 0$. This renormalization means the change in the sheet current $J_M=\tilde H_t(0)$ and leads to an additional term proportional to $\lambda$ in the magnetic moment of the sample (see Appendix \ref{B} where the elliptic cylinder in the transverse magnetic field is considered). However, $\partial H_t(t)/\partial t|_{t=0}=0$ in the case of the slab, which has the flat faces,  and the additional term does not appear in Eq.~(\ref{15}).

Superconducting materials are usually anisotropic, and their anisotropy is characterized by the parameter $\varepsilon$, which is the ratio of the London penetration depths $\lambda$ and $\lambda_{\parallel}$ for the currents flowing perpendicular to the anisotropy axis of the material and parallel to it, respectively \cite{bl}. Formula (\ref{15}) does not depend on  $\varepsilon$ and contains only the London penetration depth $\lambda$  when the direction of the magnetic field (the $y$ axis) coincides with the anisotropy axis. (In this case, the currents flow in the plane perpendicular to this axis.) If the anisotropy axis coincides with the $x$ axis,  the Meissner currents $j_z$ are still perpendicular to it. Only near the edges of the sample, the direction of the currents coincides with the axis, and in Eq.~(\ref{15}), we should use the appropriate London penetration depth $\lambda_{\parallel}$, replacing the terms $(L-2\lambda)$ and $2\lambda/L$ by  $(L-2\lambda_{\parallel})$ and $2\lambda_{\parallel}/L$, respectively.

\subsection{Comparison of the results}\label{IIIB}

 \begin{figure}[t] % %%%%%%%%%%%%%%%%%%%%%%%%%%%%%%%%%%%%%
 \centering  \vspace{+9 pt}
\includegraphics[scale=.47]{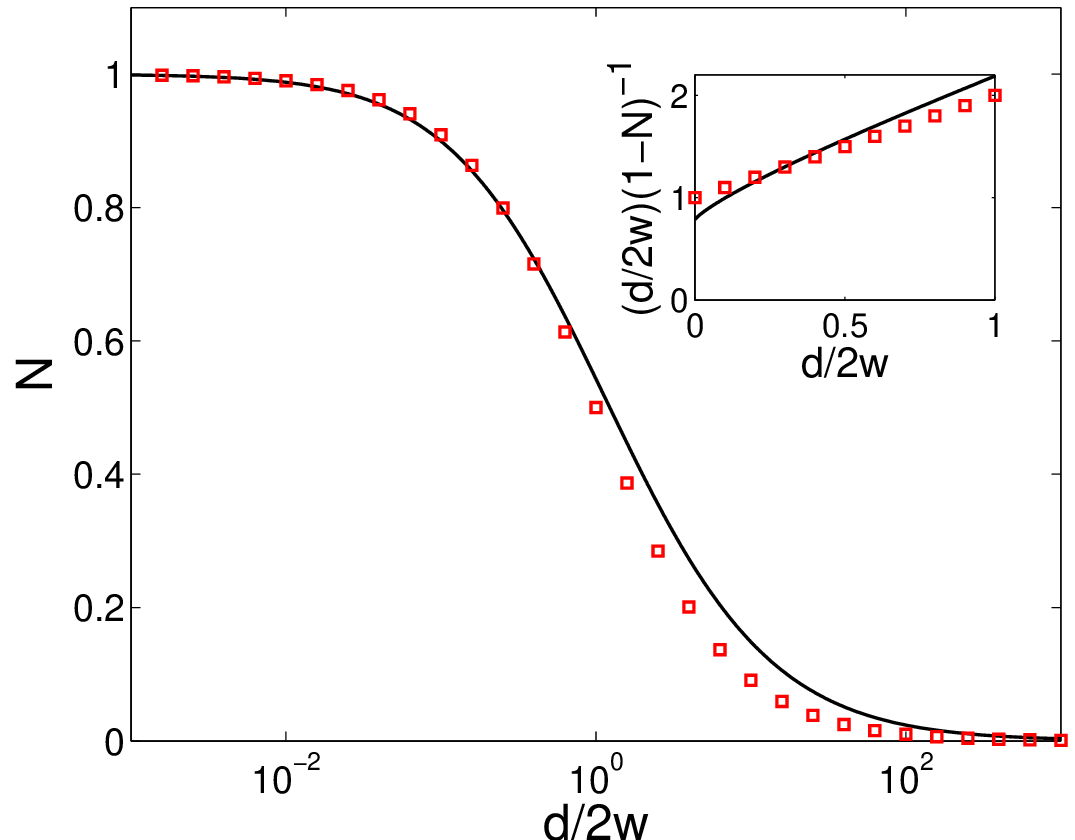}
\caption{\label{fig2} Dependence of the effective demagnetization factor $N$, Eq.~(\ref{18}), on the aspect ratio $d/2w$ of the slab (black solid line).  For comparison, the dependence described by expression (\ref{21}) is also presented (red squares). Inset: Behavior of $(1-N)^{-1}$ in the region of small $d/2w$. The solid line and the squares correspond to Eqs.~(\ref{18}) and (\ref{21}), respectively.  }
\end{figure}   %%%%%%%%%%%%%%%%%%%%%%%%%%%%%%%%%%%%%%%%%%

Let us compare the obtained results with the formula presented by Prozorov et al.\ \cite{proz00} for the 2D case ($L\to \infty$) corresponding to the discussed situation:
\begin{eqnarray}\label{20}
 m_y(\lambda)=
 -\frac{H_aV}{1-N}\left[1-\frac{\lambda}{R}\right],
 \end{eqnarray}
where  $N$ and $R$ are the effective demagnetization factor and the effective size of the sample, respectively. Using results of numerical calculations, the following approximate expressions were proposed for these $N$ and $R$ \cite{proz00}:
\begin{eqnarray}\label{21}
 \frac{1}{1-N}&\approx& 1+\frac{2w}{d}, \\
 R&\approx& \frac{w}{1+\arcsin(1/a)}, \label{22}
 \end{eqnarray}
where $a^2=1+(d/w)^2$. In Fig.~\ref{fig2}, the quantities $N$ described by formulas (\ref{18}) and (\ref{21}) are compared. For the plate in the parallel magnetic field (when $d\gg w$ and $m\to 1$), expression (\ref{18}) gives $N=0$, since $f(1,1)=1$ and $f(1,1-m)\approx \pi (1-m)/4$ in this limiting case. Formula (\ref{21}) leads to the same value. In the opposite case of the thin strip in the perpendicular magnetic field (when $d\ll w$ and $m\ll 1$), formula (\ref{21}) yields $(1-N)\approx d/2w$. This $(1-N)$ is comparable with the value, $1-N\approx m\approx 2d/\pi w\approx 0.64d/w$, that follows from expression (\ref{18}). However since the magnetic moment $m_y\propto (1-N)^{-1}\propto w/d \gg 1$, the small difference in $N$ is not negligible.

 \begin{figure}[t] % %%%%%%%%%%%%%%%%%%%%%%%%%%%%%%%%%%%%%
 \centering  \vspace{+9 pt}
\includegraphics[scale=.47]{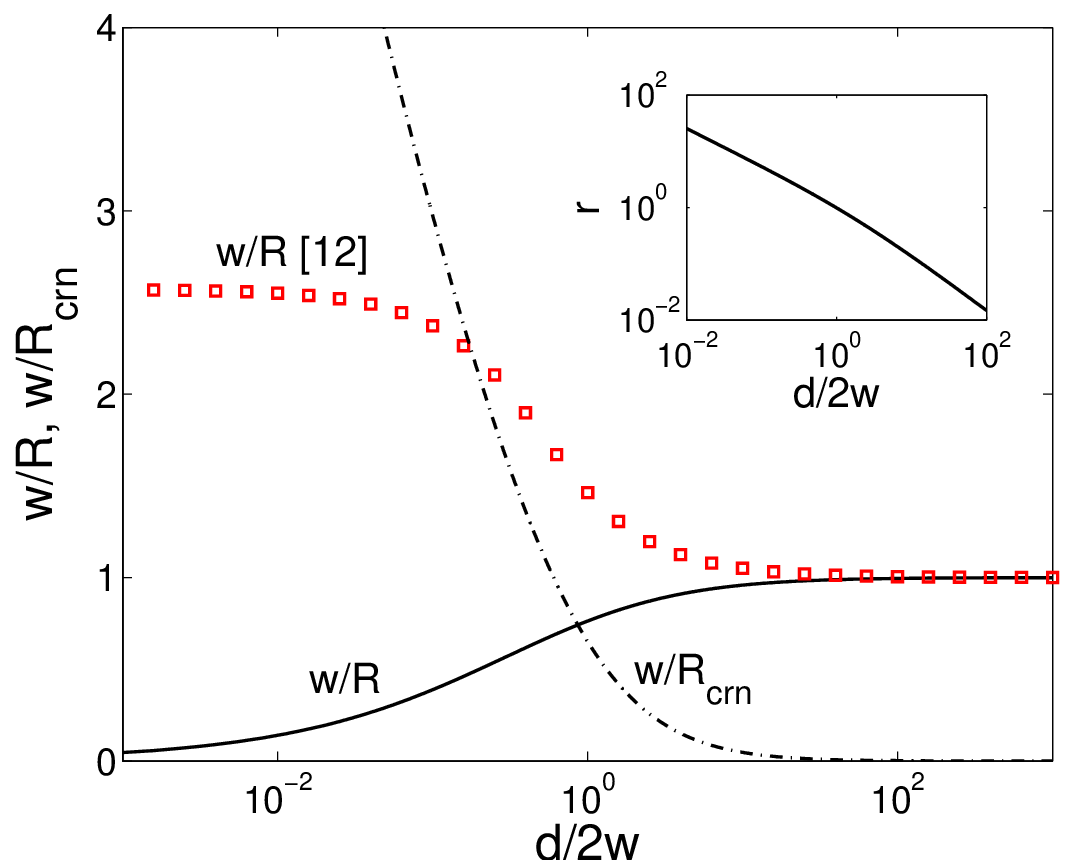}
\caption{\label{fig3} Dependences of the quantities $w/R$ (solid line) and $w/R_{\rm crn}$ (dash-and-dot line) on the aspect ratio $d/2w$ of the slab. Here $R$ and $R_{\rm crn}$ are the two effective sizes that are calculated with formulas (\ref{16}) and (\ref{17}), respectively.  For comparison, the $d/2w$ dependence of $w/R$ proposed in Ref.~\cite{proz00}, Eq.~(\ref{22}), is also presented (red squares). Inset: The dependence of the parameter $r$ defined by formula (\ref{24}) on $d/2w$. }
\end{figure}   %%%%%%%%%%%%%%%%%%%%%%%%%%%%%%%%%%%%%%%%%%

In the limit $d\gg w$, expressions (\ref{16}) and (\ref{22}) lead to one and the same value $R=w$. In the opposite case of the thin strips, Eq.~(\ref{16}) yields
 \[
R\approx  \frac{\pi w}{4\sqrt{m}}\approx \frac{\pi w}{4}\left(\frac{\pi w}{2d}\right)^{1/2},
 \]
whereas formula (\ref{22}) gives a much smaller value of the effective size, $R\approx w/[1+(\pi/2)]\approx 0.39w$ (Fig.~\ref{fig3}). Moreover, it was noticed in the recent paper \cite{proz_pr-appl21}, in which the effective size $R$ was numerically calculated for the superconducting cylinder of the length $2c$ and of the radius $a$ in the longitudinal magnetic field, that this $R$ tends to zero with decreasing $c/a$. In other words, the saturation of $R$ predicted by Eq.~(\ref{22}) at small aspect ratio does not seem to occur, and the discrepancy between $R$ derived from Eq.~(\ref{16}) and  $R$ numerically calculated may be even larger than the deviation following from Eq.~(\ref{22}).

However, the most important feature of formula (\ref{15}) is that it
contains the nonanalytic-in-$\lambda$ term $(\lambda/R_{\rm crn})^{2/3}$ which is absent in Eq.~(\ref{20}).
In contrast to $R$, the ratio $R_{\rm crn}/w$ decreases with decreasing the aspect ratio $d/2w$ of the sample (Fig.~\ref{fig3}), and at $d\ll w$, we have
 \[
R_{\rm crn}\approx w\frac{\sqrt{6}\pi^{3/2}m^{1/4}}{32} \approx \frac{\pi w}{16}\left(\frac{9\pi d}{2w}\right)^{1/4}.
 \]
Therefore, the relative role of the two terms in Eq.~(\ref{15}) can be different for thin and thick samples (see below).
Moreover, if the cross section of a real sample has rounded corners, and the radius $\rho$ of the curvature for these corners significantly exceeds $\lambda$, the nonanalytic term disappears. This means that if the sharp corners of the slab are damaged, the magnitude of the nonanalytic term may noticeably  decrease, and relative role of the two terms in the magnetic moment can be different for real samples of the same dimensions.

It is also necessary to keep in mind  that the nonanalytic-in-$\lambda$ term defined by $R_{\rm crn}$ may contain an additional numerical factor $c_0$ of the order of unity since the exact distribution of $j_z(x,y)$ near the corners is unknown. [We have replaced this distribution by the constant $j_{\rm crn}$ in  the calculation of the moment $M_{\rm crn}$ corresponding to the corner region (iii), see Sec.~\ref{IIIA}]. Nevertheless, it is shown  in Appendix \ref{C} that this cut-off of the current has no effect on the power $2/3$ of $\lambda$ in Eq.~(\ref{15}), and  the nonanalytic-in-$\lambda$ term in the magnetic moment is inherent in any superconducting cylinder in the transverse magnetic field if its cross section contains a sharp corner characterized by the condition $\rho\ll \lambda$. On the other hand, the rounded corners with $\rho> \lambda$ give a large linear-in-$\lambda$ contribution to the magnetic moment, and at $\rho\sim \lambda$, this contribution transforms into the nonanalytic term (Appendix \ref{C}).

Existence of the term $(\lambda/R_{\rm crn})^{2/3}$ in the magnetic moment can explain the discrepancy between formula (\ref{16}) and the results of the numerical calculation of $R$ in Refs.~\cite{proz00,proz_pr-appl21}. In these numerical calculations, the correction to the magnetic moment was approximated by the linear-in-$\lambda$ term only, and so the obtained effective size of the samples was affected by the term proportional to $\lambda^{2/3}$. As a result, the $R$ obtained  numerically \cite{proz_pr-appl21} demonstrates the behavior characteristic of the nonanalytic term in the region of its dominance (i.e., at $d/2w\ll 1$).

\section{The London penetration depth} \label{IV}

Below we use the designation $\Delta Q\equiv Q(T)-Q(T_{\rm min})$ where $Q$ is any quantity dependent on the temperature $T$. Then,
with equation (\ref{15}), a generalization of formula (\ref{1}) yields the following shift of the  frequency in the tunnel diode oscillator technique applied to the slab with sharp edges:
\begin{eqnarray}\label{23}
\frac{\Delta f}{\delta f}=-\frac{\Delta\lambda}{R} -\frac{\Delta (c_0 \lambda^{2/3})}{R_{\rm crn}^{2/3}}-\frac{2\Delta\lambda}{L},
\end{eqnarray}
where we have included an additional  numerical factor $c_0\sim 1$ in the nonanalytic-in-$\lambda$ term (see the previous section). Note that this constant $c_0$ is independent of the aspect ratio of the slab and is determined by the angles of its cross-section corners (i.e., for all rectangular cross sections, $c_0$ has a universal value). In the case of a long slab ($L\gg d, w$), which is considered here, the last term in formula (\ref{23}) may be neglected.

In the limiting case $\Delta\lambda\ll \lambda$, one has $\Delta(c_0\lambda^{2/3})\propto \Delta\lambda$, and the total shift in the frequency is still proportional to $\Delta\lambda$, but the coefficient before $\Delta\lambda$ now depends on the temperature.   In this limiting case, it is easy to estimate the relative role of the first two terms in formula (\ref{23}), analyzing their ratio,
\[
    \frac{2}{3}c_0\left(\frac{w}{\lambda}\right)^{1/3}r(m),
\]
where the factor $r(m)$,
 \begin{eqnarray}\label{24}
 r(m)=\!\frac{R}{w^{1/3}R_{\rm crn}^{2/3}}=\! \left(\frac{8(1-m)[f(1,1-m)]^{2}}{3m^2}\right)^{1/3}\!\!\!\!\!\!,
 \end{eqnarray}
depends only on the aspect ratio of the slab (Fig.~\ref{fig3}).
For thin strips ($m\approx 2d/\pi w\ll 1$), the parameter $r$ is approximately equal to $1.87(w/d)^{2/3}\gg 1$. This means that the term $\Delta\lambda/R$ in Eq.~(\ref{23}) can be neglected. On the other hand, for $d/2w\gg 1$ (i.e., for a plate in the magnetic field parallel to its surface),  the factor $r$  is less than unity,  $r\approx 3w/d\ll 1$. In this situation, the term $\Delta\lambda/R$ can prevail over $\Delta(c_0\lambda^{2/3})/R_{\rm crn}^{2/3}$. The boundary value of $d/2w$, at which the relative role of the terms in Eq.~(\ref{23}) changes, is determined by the condition $r(m)\sim (3/2c_0)(\lambda/w)^{1/3}$. For example, if $\lambda= 0.14$\,$\mu$m, and $w=50$\,$\mu$m [i.e., if $(\lambda/w)^{1/3} \approx 1/7$], the appropriate $r(m)\sim 0.2$, and the boundary value of $d/2w$ is approximately equal to $6$. However, since the nonanalytic term can noticeably decrease in real samples, we conclude that  both the terms in formula (\ref{23}) can have comparable values for $d\sim 2w$.

 \begin{figure}[t] % %%%%%%%%%%%%%%%%%%%%%%%%%%%%%%%%%%%%%
 \centering  \vspace{+9 pt}
\includegraphics[scale=.46]{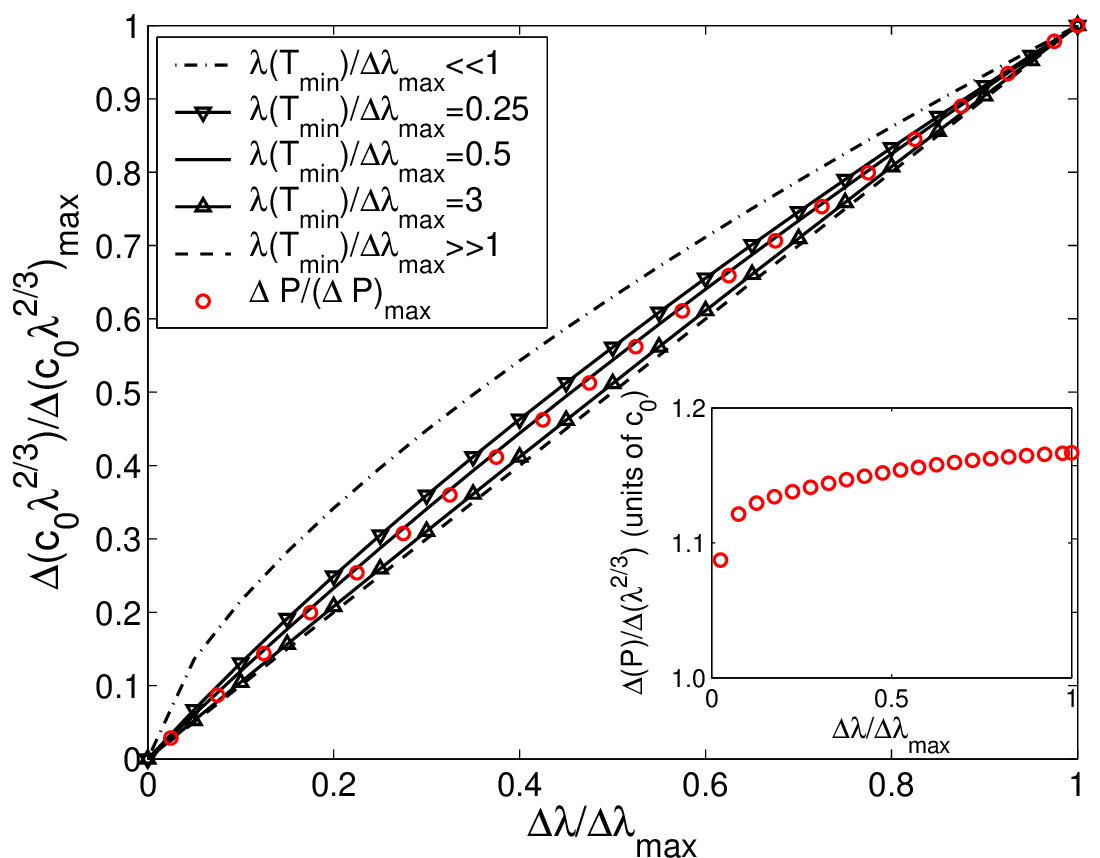}
\caption{\label{fig4} Dependence of the ratio $\Delta(c_0\lambda^{2/3})/\Delta(c_0\lambda^{2/3})_{\rm max}$ on $\Delta\lambda/\Delta\lambda_{\rm max}$ (black lines) for the slab with sharp edges for the several values of the parameter $\lambda(T_{\rm min})/\Delta\lambda_{\rm max}$. Here $\Delta\lambda_{\rm max}$ is maximal value of $\Delta\lambda(T)$ obtained at $T=T_{\rm max}$, and $\Delta(c_0\lambda^{2/3})_{\rm max}$ is the value of $\Delta(c_0\lambda^{2/3})$ at the same temperature.
The red circles show $\Delta(P)/\Delta(P)_{\rm max}$ where $\Delta(P)$ is defined by formulas (\ref{27}) and (\ref{29}), $\Delta(P)_{\rm max}$ is $\Delta(P)$ at $T=T_{\rm max}$, $\lambda(T_{\rm min})/ \Delta\lambda_{\rm max}=0.25$, $c_1=c_0$, $\rho/\lambda(T_{\rm min})=10$, $\nu_0=0.5$.
Inset: The ratio $\Delta(P)/\Delta(\lambda^{2/3})$ versus $\Delta\lambda/\Delta\lambda_{\rm max}$  where $\Delta(P)$ correspond to the red circles in the main panel, whereas $\Delta(\lambda^{2/3})$ is calculated with the false value of $\lambda(T_{\rm min})/\Delta\lambda_{\rm max}=0.5$ (see text). }
\end{figure}   %%%%%%%%%%%%%%%%%%%%%%%%%%%%%%%%%%%%%%%%%%

All the formulas of this paper have been derived under the assumption that the magnetic field is directed along the thickness $d$ of the slab (i.e., along the $y$ axis). However, if the magnetic field is directed along its width $2w$, we still can use all the obtained formulas, replacing $d$ by $2w$ and $2w$ by $d$ in them. In this situation,  the new parameter $m$ is equal to the value of $1-m$ for the case when $H$ is parallel to the $y$ axis.

Consider now the long slab for which $d$ is not equal to $2w$ (i.e., $m$ does not coincide with $1/2$). Let the frequency shifts $(\Delta f/\delta f)_x$ and $(\Delta f/\delta f)_y$  be measured in this sample for the magnetic fields directed along the $x$ and $y$ axes, respectively. Since the effective sizes $R$, $R_{\rm crn}$  for $H_a\parallel y$ and $R'$, $R_{\rm crn}'$ for $H_a\parallel x$  can be calculated with formulas (\ref{16}) and (\ref{17}), these shifts enable one to find the quantities $\Delta\lambda$ and $\Delta(c_0\lambda^{2/3})$, using the two equations (\ref{23}) written for these two directions of the magnetic field,
\begin{eqnarray}\label{25}
\left(\frac{\Delta f}{\delta f}\right)_y=-\frac{\Delta\lambda}{R} -\frac{\Delta (c_0\lambda^{2/3})}{R_{\rm crn}^{2/3}}, \nonumber \\
\left(\frac{\Delta f}{\delta f}\right)_x=-\frac{\Delta\lambda}{R'} -\frac{\Delta (c_0\lambda^{2/3})}{[R_{\rm crn}']^{2/3}}.
\end{eqnarray}
Condition $d\neq 2w$ provides a nonzero determinant for these two linear equations. Thus, the measurements of $(\Delta f/\delta f)_x$ and $(\Delta f/\delta f)_y$  make it possible to find the change in the London penetration depth, $\Delta\lambda(T)$, and to avoid a  ``contamination'' of this quantity by the contribution which results from the edges of the slab and which is difficult to control experimentally.

However, an analysis of the contribution $\Delta(c_0\lambda^{2/3})$ can open up additional possibilities for the determination of $\lambda$. The quantity $\Delta(\lambda^{2/3})$ can be written as follows:
  \[
  \frac{\Delta(\lambda^{2/3})}{(\Delta\lambda_{\rm max})^{2/3}}=\!\left(\left[\frac{\lambda(T_{\rm min})}{\Delta\lambda_{\rm max}}+ \frac{\Delta\lambda}{\Delta\lambda_{\rm max}}\right]^{2/3}\!\!\!\!\!\!\!-\!\left[\frac{\lambda(T_{\rm min})}{\Delta\lambda_{\rm max}}\right]^{2/3}\right)
   \]
where  $\Delta\lambda_{\rm max}$ is the maximal value of $\Delta\lambda$ obtained at some temperature $T_{\rm max}$. If the ratio of the term $\Delta(c_0\lambda^{2/3})$ to its value $\Delta(c_0\lambda^{2/3})_{\rm max}$ at this $T_{\rm max}$ is plotted as a function of $\Delta\lambda/\Delta\lambda_{\rm max}$, this function depends on the only parameter $\lambda(T_{\rm min})/\Delta\lambda_{\rm max}$ (Fig.~\ref{fig4}). According to this figure, if $\lambda(T_{\rm min})/\Delta\lambda_{\rm max}\ge 3$, the function practically coincides with the linear dependence. This means that the term $\Delta(c_0\lambda^{2/3})$ is almost proportional to $\Delta\lambda$ in this interval of $\Delta\lambda$, and  the coefficient before $\Delta\lambda$ is equal to
  \begin{eqnarray}\label{26}
  \frac{2c_0}{3[\lambda(T_{\rm min})]^{1/3}}.
  \end{eqnarray}
If this coefficient is found experimentally, it gives the estimate of $\lambda(T_{\rm min})$  since $c_0\sim 1$.

If $\Delta\lambda$ and $\Delta(c_0\lambda^{2/3})$ are found in a sufficiently wide temperature interval, $\Delta\lambda_{\rm max} \gtrsim \lambda(T_{\rm min})$, the dependence of $\Delta(c_0\lambda^{2/3})$ on $\Delta\lambda$ becomes nonlinear (Fig.~\ref{fig4}), and this nonlinearity is determined by the  parameter $\lambda(T_{\rm min})/\Delta\lambda_{\rm max}$. It is important that the shape of this dependence has the universal form for samples characterized by the condition $\rho \ll\lambda(T_{\rm min})$. Therefore, the measurements of $(\Delta f/\delta f)_x$ and $(\Delta f/\delta f)_y$ in the wide temperature interval provide possibility not only to find the change of the London penetration depth, $\Delta\lambda(T)$, but also to estimate the value of $\lambda(T_{\rm min})/\Delta\lambda_{\rm max}$ [and hence, the value of $\lambda(T_{\rm min})$]  for the slab with the sharp edges.
Such results could complements the information on $\lambda(T_{\rm min})$  obtained by other methods, e.g., using the superconductive coating \cite{proz-apl}.

Knowing $\lambda(T_{\rm min})$ and $\Delta\lambda(T)$, one can calculate $\Delta(\lambda^{2/3})$. Then, for the slab with sharp edges,  the ratio $\Delta(c_0\lambda^{2/3})/\Delta(\lambda^{2/3})$ must be independent of the temperature and is equal to the constant $c_0$. This makes it possible to find $c_0$ experimentally and to verify the sharpness of the edges of the sample.

\subsection{Slab with rounded edges}

Above we have analyzed the case of the slab with the sharp edges when $\rho\ll\lambda(T_{\rm min})$. Discuss now the situation when a part of the edges is damaged,  and these edges have rounded corners with the averaged  radius of the curvature $\rho> \lambda(T_{\rm min})$. Let the ratio of their length to the total length $4L$ of the edges in the slab be equal to $\nu$. This parameter $\nu$ characterizes the quality of the edges in a real sample. Taking into account formula (\ref{c3}), we may consider a simple model in which the term $\Delta(c_0\lambda^{2/3})$ in formula (\ref{23}) and in equations (\ref{25}) is replaced by the expression,
\begin{eqnarray}\label{27}
 \Delta\left(c_0\lambda^{2/3}(1-\nu)+ c_1\nu\frac{\lambda}{\rho^{1/3}}\right) \equiv \Delta\big(P(\lambda)\big),
 \end{eqnarray}
where $c_1$ is a number of the order of unity. We take $c_1=c_0$ below. In this case, the quantity $P$ transforms into $c_0\lambda^{2/3}$ at $\lambda=\rho$. It is seen from formula (\ref{27}) that the factor before $\lambda^{2/3}$ decreases as compared to its value $c_0$ for the slab with sharp edges, whereas the linear-in-$\lambda$ contribution to the frequency shift increases, and $\Delta\lambda/R$ in Eq.~(\ref{23}) is replaced by $\Delta\lambda/\tilde R$ where $\tilde R$ is renormalized size,
 \begin{eqnarray}\label{28}
  \frac{1}{\tilde R}=\frac{1}{R}+c_1\nu\frac{1}{\rho^{1/3}R_{\rm crn}^{2/3}}.
 \end{eqnarray}
In other words, for such samples, the two contributions to the frequency shift are affected by the parameters $\nu$ and $\rho$, which can hardly be controlled in experiments. For this reason, it is still convenient to measure the frequency shifts for the two direction of the magnetic field and to solve the set of the equations similar to  set (\ref{25}). (In this set, the variables are $\Delta\lambda$ and $\Delta P$.) This procedure enables one to separate the changes in the London penetration depth, $\Delta\lambda$, and in the quantity $P$,  $\Delta\big(P(\lambda)\big)$, from each other. Note that $\Delta\lambda$ thus obtained is not affected by the nonzero $\nu$ and $\rho$. However, if the ratio of the term $\Delta(P)$ to its value $\Delta(P)_{\rm max}$ at the temperature $T_{\rm max}$ is plotted as a function of $\Delta\lambda/ \Delta\lambda_{\rm max}$, this function depends not only on the parameter $\lambda(T_{\rm min})/\Delta\lambda_{\rm max}$ but also on $\nu$ and $\rho$. Moreover, the parameter $\nu$ itself depends on $\rho/\lambda(T)$. Indeed, when $T$ increases, this ratio decreases, and $\nu\to 0$ if $\lambda(T)$ essentially exceeds $\rho$.

In Fig.~\ref{fig4}, we demonstrate the plot of
$\Delta(P)/\Delta(P)_{\rm max}$ versus $\Delta\lambda/ \Delta\lambda_{\rm max}$, using the simplest model for the dependence of $\nu$ on $\rho/\lambda$,
\begin{eqnarray}\label{29}
  \nu(\rho/\lambda)=\nu_0[1-\tanh(\lambda/\rho)],
  \end{eqnarray}
which correctly reproduces the values of $\nu$ at small and large $\lambda/\rho$ ($\nu_0$ is a constant). Interestingly, this plot calculated for $\lambda(T_{\rm min})/\Delta\lambda_{\rm max}=0.25$, $\nu_0=0.5$, $\rho/\lambda(T_{\rm min})=10$ is close to $\Delta(c_0\lambda^{2/3})/\Delta(c_0\lambda^{2/3})_{\rm max}$ plotted for the slab with $\rho/\lambda(T_{\rm min})\ll 1$ and another $\lambda(T_{\rm min})/ \Delta\lambda_{\rm max}=0.5$. Thus, a sample with the rounded edges can ``imitate'' the slab with the sharp corners, and this can lead to the incorrect estimate of the absolute value of $\lambda(T_{\rm min})$. However, the ratio $\Delta(P)/\Delta(\lambda^{2/3})$, where $\Delta(\lambda^{2/3})$ is calculated with the estimated $\lambda(T_{\rm min})$  and the found dependence $\Delta\lambda(T)$, is not constant for the sample with rounded edges and differs from $c_0$ (inset in Fig.~\ref{fig4}). This variation of the ratio may serve as an indication of the presence of the rounded edges in the sample.

\section{Conclusions}\label{V}

It is shown that for a long slab with rectangular cross section, the temperature shift of the resonant frequency in the tunnel diode oscillation technique  comprises the two parts. The first part is proportional to the  change $\Delta\lambda$ of the London penetration depth with the temperature. This part exists for any superconductor with the smooth surface (e.g. for the long elliptic cylinder in the transverse magnetic field), and it was discussed in numerous publications. In this paper, the explicit expressions for this part are derived in the cases of the slab and the elliptic cylinder. The second part of the frequency shift is due to the sharp edges of the slab and is proportional to the change of $\lambda^{2/3}$ with increasing temperature. This part can reach a large value especially for thin samples. Measurements of the frequency shifts for the magnetic fields applied along the thickness and the width of the slab give  possibility to separate both the parts of the shift from each other. If these parts are found for the slab with sharp edges, they enable one not only to determine  the temperature dependence of the London penetration depth, but also to estimate its absolute value.

\appendix

\section{Calculation of magnetic moment}\label{A}

Let us prove relation (\ref{14}). Consider the following transformations of the integral:
\begin{eqnarray}\label{a1}
\int zj_xdxdydz&=&\int z\left(j_x x|^{x=w}_{x=-w} -\int x\frac{\partial j_x}{\partial x}dx\right)dydz \nonumber\\
&=&\int z dydz\int x(\frac{\partial j_y}{\partial y} + \frac{\partial j_z}{\partial z})dx \nonumber \\
&=&\int\!\!xz j_y|^{y=d/2}_{y=-d/2}dxdz+\!\!\int\!\!xz\frac{\partial j_z}{\partial z} dxdydz \nonumber \\
&=&\int x\left(j_z z|^{z=L/2}_{z=-L/2} -\int j_zdz\right)dxdy \nonumber \\
&=&-\int xj_z dxdydz ,
  \end{eqnarray}
where we have taken into account that ${\rm div}{\bf j}=0$, and at the surfaces of the parallelepiped, one has  $j_x|_{x=\pm w}=j_y|_{y=\pm d/2}=j_z|_{z=\pm L/2}=0$. With equality  (\ref{a1}), we arrive at relation (\ref{14}),
\begin{eqnarray}\label{a2}
M_y=\frac{1}{2}\int (zj_x-xj_z)dxdydz=-\int xj_z dxdydz.~~~
 \end{eqnarray}

It is necessary to emphasize that this formula is valid not only for the parallelepiped-shaped sample but also for any three-dimensional body. Indeed, consider a parallelepiped $U$ that includes the body and is larger than it. Since the currents vanish outside the body, we may carry out the integration over  $U$  in equation (\ref{a1}). Then, all the transformations in this equation remain true in spite of the jumps of the currents occurring inside the region of the integration.

\section{Elliptic cylinder in transverse magnetic field}\label{B}

\begin{figure}[t] % %%%%%%%%%%%%%%%%%%%%%%%%%%%%%%%%%%%%%
 \centering  \vspace{+9 pt}
\includegraphics[scale=.48]{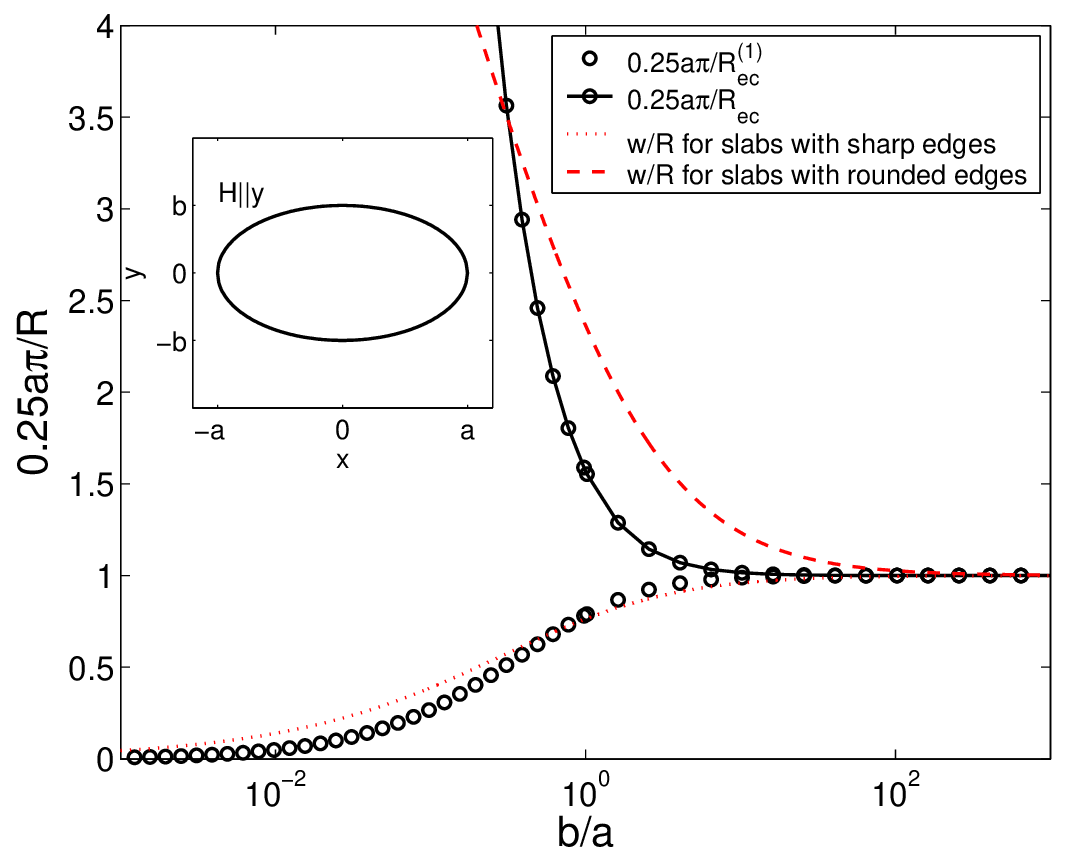}
\caption{\label{fig5}
The dependence of the effective size $R_{ec}$ of the elliptic cylinder in the transverse magnetic field $H\parallel b$ on the ratio $b/a$ where $2b$ and $2a$ are the axes of its elliptic cross section. The solid line with circles shows the ratio $a\pi/(4R_{\rm ec})$, Eqs.~(\ref{b12}) and (\ref{b13}), whereas the circles correspond to the contribution $a\pi/(4R^{(1)}_{\rm ec})$ to this ratio, Eqs.~(\ref{b5}), (\ref{b6}). For comparison, the dependences of $w/R$ on $d/2w$ are shown for the slab with sharp edges (red dotted line), Eq.~(\ref{16}), and for the  slab with rounded edges (red dashed line). In the latter case, $R=\tilde R$, $\tilde R$ is given by formula (\ref{28}) with $\nu=1$, and we take $c_1=\pi/4$, $\rho/w=0.05$ here. Inset: The cross section of the elliptic  cylinder.
 }
\end{figure}

Consider a superconducting cylinder with an elliptical cross section. Let the axes of this cross section be $2a$ and $2b$ (inset in Fig.~\ref{fig5}). The cylinder infinitely extends in the $z$ direction, and the applied magnetic field ${\bf H}_a=(0,H_a,0)$ is directed along the $y$ axis coinciding with the $b$ axis.

Since this cylinder is the specific case of an ellipsoid, we can use the concept of the demagnetization factor for the calculation of the Meissner sheet currents in the superconductor. This sheet current flowing along the $z$ axis at some point ($x,y$) of the surface of the cylinder is equal to
\begin{eqnarray}\label{b1}
 J_M(x,y)=\frac{H_a}{1-N}n_x.
\end{eqnarray}
Here $N=a/(a+b)$ is the demagnetization factor of the cylinder \cite{osborn},  $n_x$ is the $x$ component of the outward normal to its surface at the point ($x,y$),
\begin{eqnarray}\label{b2}
 n_x=\frac{b\cos\phi}{\sqrt{a^2\sin^2\phi+b^2\cos^2\phi}},
\end{eqnarray}
and the angle $\phi$ defines the position of the point on the surface, $x=a\cos\phi$, $y=b\sin\phi$.

The magnetic moment per the unit length, $M_y$, is calculated as follows:
\begin{eqnarray}\label{b3}
M_y&=&-\!\!\int\!\!j_zxdxdy=-\!\int\!\!\frac{ J_Mdl}{\lambda}\!\!\int_{-\infty}^0\!\!\!\!\!\!\exp(\frac{t}{\lambda}) [x(l)+t n_x]dt\nonumber \\
&=&-\int\!\![x(l)-\lambda n_x]J_Mdl,
\end{eqnarray}
where the integration over the cross section of the cylinder is replaced by the integration along the ellipse (coordinate $l$) and along $t$, the local coordinate perpendicular to the ellipse at its point ($x(l),y(l)$). Since $\lambda$ is small, the integration of $j_z=(J_M/\lambda)\exp(t/\lambda)$ over $t$ can be extended to $-\infty$ ($t=0$ on the ellipse). Note that formula (\ref{b3}) is  applicable to a cylinder with an arbitrary cross section, the boundary of which has no sharp corners. (In this case, $l$ is the coordinate along the boundary.)

Inserting formulas (\ref{b1}) and (\ref{b2}) into expression (\ref{b3}) and taking into account that
\[
dl\!=\!d\phi\sqrt{a^2\sin^2\!\phi\!+\!b^2\cos^2\!\phi},
\]
we arrive at
\begin{eqnarray}\label{b4}
M_y= -\frac{H_aS}{1-N}\left(1-\frac{\lambda}{R^{(1)}_{\rm ec}}\right).
\end{eqnarray}
Here $S=\pi ab$ is the area of the ellipse, and for $b\le a$, the effective size, $R^{(1)}_{\rm ec}$, is given by
\begin{eqnarray}\label{b5}
R^{(1)}_{\rm ec}=\frac{\pi a}{4}\frac{k^2}{\sqrt{1-k^2}(K(k)-E(k))},
\end{eqnarray}
where $k=\sqrt{1-(b/a)^2}$, and $K(k)$, $E(k)$ are
the complete elliptic integrals of the first and second kinds, respectively. If $b\ge a$, we have
\begin{eqnarray}\label{b6}
R^{(1)}_{\rm ec}=\frac{\pi a}{4}\frac{k^2}{[E(k)-(1-k^2)K(k)]},
\end{eqnarray}
where $k=\sqrt{1-(a/b)^2}$.

For the elliptic cylinder, its surface has a curvature, and the renormalization of the surface magnetic field, Eq.~(\ref{19}), becomes important. In the limit $\lambda\to 0$, the solution of the Laplace equation for the magnetic field outside the superconducting elliptic cylinder gives,
\begin{eqnarray}\label{b7}
\frac{1}{H_t(0)}\frac{\partial H_t}{\partial t}|_{t=0} =-\frac{ab}{(a^2\sin^2\phi+b^2\cos^2\phi)^{3/2}}.
\end{eqnarray}
The nonzero $(\partial H_t/\partial t)_{t=0}$ means that an increase of $\lambda$ decreases the sheet currents $J_M$, and this change leads to another correction to the magnetic moment  $M_y$. This correction is also proportional to $\lambda$,
\begin{eqnarray}\label{b8}
M_y= -\frac{H_aS}{1-N}\left(1-\frac{\lambda}{R^{(2)}_{\rm ec}}\right),
\end{eqnarray}
where $R^{(2)}_{\rm ec}$ at $b\le a$ is given by
\begin{eqnarray}\label{b9}
R^{(2)}_{\rm ec}=\frac{\pi b}{4}\frac{k^2}{E(k)-(1-k^2)K(k)},
\end{eqnarray}
and $k=\sqrt{1-(b/a)^2}$. If $b\ge a$, we have
\begin{eqnarray}\label{b10}
R^{(2)}_{\rm ec}=\frac{\pi b}{4}\frac{k^2}{\sqrt{1-k^2}[K(k)-E(k)]},
\end{eqnarray}
where $k=\sqrt{1-(a/b)^2}$. The sum of the terms (\ref{b4}) and (\ref{b8}) leads to the linear-in-$\lambda$ contribution  to $M_y$,
\begin{eqnarray}\label{b11}
M_y= -\frac{H_aS}{1-N}\left(1-\frac{\lambda}{R_{\rm ec}}\right),
\end{eqnarray}
with $1/R_{\rm ec}=(1/R^{(1)}_{\rm ec})+(1/R^{(2)}_{\rm ec})$,
\begin{eqnarray}\label{b12}
R_{\rm ec}=\frac{\pi b}{4E(k)},\ \ \ k=\sqrt{1-(b/a)^2}, \\
R_{\rm ec}=\frac{\pi a}{4E(k)},\ \ \ k=\sqrt{1-(a/b)^2}, \label{b13}
\end{eqnarray}
where formulas (\ref{b12}) and (\ref{b13}) correspond to the cases $b\le a$ and $b\ge a$, respectively.

In Fig.~\ref{fig5}, we show the dependences of $\pi a/4R^{(1)}_{\rm ec}$ and of $\pi a/4R_{\rm ec}$ on the ratio $b/a$ of the axes of the cylinder. The dependence of $\pi a/4R^{(1)}_{\rm ec}$  is similar to the dependence of $w/R$ for the slab with sharp edges on its aspect ratio $d/2w$. This similarity stems from the fact that both $R^{(1)}_{\rm ec}$ and $R$ are associated with the redistribution of the Meissner currents flowing near the surface of a sample when  $\lambda$ increases. The term $\pi a/4R^{(2)}_{\rm ec}$, i.e, the difference between $\pi a/4R_{\rm ec}$ and $\pi a/4R^{(1)}_{\rm ec}$, is caused by the renormalization of the magnitude of the surface sheet currents and is due to  the curvature of the surface. A similar term exists for the slab with rounded edges. Although the region of the rounded edges is relatively small as compared to the total surface of the slab, but the curvature is large in this region. For this reason, these terms for the cylinder and for the  slab give comparable corrections to the magnetic moment.

    %%%%%%%%%%%%%%%%%%%%%%%%%%%%%%%%%%%%%%%%%%

\section{Contribution of a sharp edge to the magnetic moment }\label{C}

 \begin{figure}[t] % %%%%%%%%%%%%%%%%%%%%%%%%%%%%%%%%%%%%%
 \centering  \vspace{+9 pt}
\includegraphics[scale=.47]{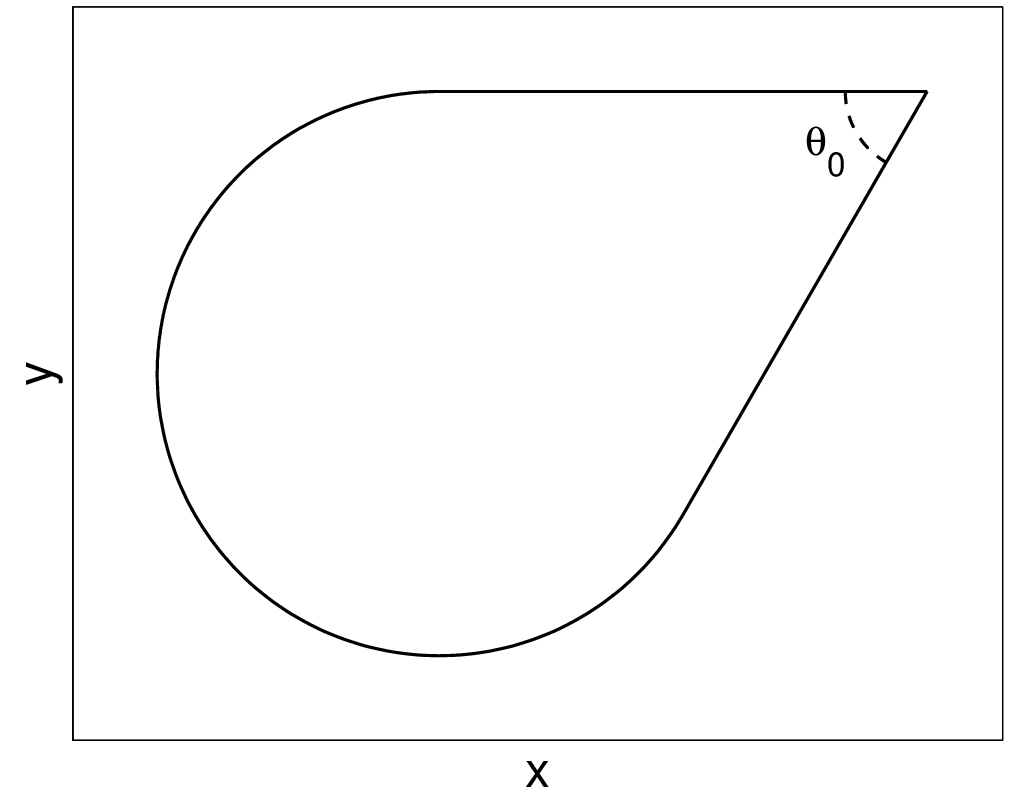}
\caption{\label{fig6} The cross section of the cylinder with an edge extending along the $z$ axis. The $\theta_0$ is the dihedral angle of this edge. }
\end{figure}   %%%%%%%%%%%%%%%%%%%%%%%%%%%%%%%%%%%%%%%%%%

Let a superconductor be a long cylinder with an arbitrary cross section in the $x-y$ plane, and let the sample have a sharp edge extending along the $z$ axis, Fig.~\ref{fig6}. The term ``sharp'' means that near the edge, the radius of curvature of the cross-section boundary is less than the London penetration depth $\lambda$. Besides, it is assumed that $\lambda\ll R_{\rm ch}$ where $R_{\rm ch}$  is the characteristic size of the cross section. The magnetic field $H_a$ is perpendicular to the $z$ axis (e.g., it is along the $y$ axis). In the vicinity of the edge, the vector potential $A_z$ outside the sample is described by the two-dimensional Laplace equation. When the sample is in the Meissner state (i.e., when $A_z=0$ on its surface),  the solution of the equation has the form $A_z\propto H_aR_{\rm ch}(r/R_{\rm ch})^n\sin(n\theta)$ in the vicinity of the edge (see problem 3 to \S 3 of Ref.~\cite{LL}). Here $r$, $\theta$ are the  cylindrical coordinates with the origin on the edge,  $n=\pi/(2\pi-\theta_0)$, and $\theta_0$ is the dihedral angle of the edge (for the slab, $R_{\rm ch}\sim w$, $\theta_0=\pi/2$, and $n=2/3$). Then, near the edge, the magnetic field at the surface and the Meissner sheet current $J_M$ are proportional to $(1/r)\partial A_z/\partial\theta\propto H_a(r/R_{\rm ch})^{n-1}$. The cut-off of the currents (see Sec.~\ref{II}) leads to that the region  $\lambda\times\lambda\times L$ near the edge gives the contribution  $\delta m_y$,
 \[
\delta m_y\!\sim\!J_M\lambda LR_{\rm ch}\!\propto\! L\lambda H_a(\lambda/R_{\rm ch})^{n-1}\!R_{\rm ch}\!\propto\! H_aV(\lambda/R_{\rm ch})^n
  \]
to the magnetic moment of the sample. Here $L$ and $V\propto LR_{\rm ch}^2$ are the length and the volume of the superconductor, respectively, and the coefficient of the proportionality in $\delta m_y$ depends on the shape of the sample (in the case of the slab, on its aspect ratio). Therefore,  a sharp edge of sample with any dihedral angle $\theta_0<\pi$ leads to the nonanalytic $\lambda^n$ contribution  ($n<1$) to the magnetic moment.

In the calculation of the magnetic moment $M_{\rm crn}$ generated by the currents in the corner region (iii) of the slab, we have assumed that the current density in this region is constant  and is equal to $j_{\rm crn}$, see Sec.~\ref{III}. The similar cut-off of the diverging current has been used in estimating $\delta m_y$ above. Let us discuss this approximation in more detail, considering the case of the slab with the rectangular cross section. The current density $j_z$ near the corner of the slab satisfies the  London equation which in the cylindrical coordinates looks like
\begin{eqnarray}\label{c1}
\frac{1}{r}\frac{\partial}{\partial r}\left( r\frac{\partial j_z}{\partial r}\right)+ \frac{1}{r^2}\frac{\partial^2 j_z}{\partial \theta^2}-\frac{j_z}{\lambda^2}=0.
\end{eqnarray}
If we introduce the dimensionless coordinate $\tilde r=r/\lambda$,
equation (\ref{c1}) becomes independent of $\lambda$. (The  Laplace equation outside the corner and the condition of the magnetic-field continuity at its boundary remain $\lambda$-independent). Taking into account formulas (\ref{11})-(\ref{13}), the current density on the surface of the corner at $\tilde r>1$ can be represented in the form, $j_z(\tilde r)=j_{\rm crn}\tilde r^{-1/3}$. Therefore, if we also introduce the dimensionless current density $\tilde j_z\equiv j_z/j_{\rm crn}$, the exact solution for $\tilde j_z$ is independent of both $\lambda$ and $H_a$. In other words, for given $\theta_0=\pi/2$, the dimensionless current density  is described by an universal function of $\tilde r$ and $\theta$ in the corner region (iii), and the total dimensionless current flowing in this region is equal to some number $I_0$, which is of the order of unity. (In our approximation  which is based on the cut-off procedure, this dimensionless current $I_0=1$). Then, with the exact solution of the London equation, the additional numerical factor $c_0=(3-I_0)/2$ appears in the nonanalytical term $(\lambda/R_{\rm crn})^{2/3}$.

The value of this factor can be found in the numerical calculations similar to those carried out by Prozorov for  a finite cylinder in the longitudinal magnetic field \cite{proz_pr-appl21}. In Ref.~\cite{proz_pr-appl21}, the quantity $[M(1-N)/VH_a]+1$ was calculated for different $\lambda$, and the results were approximated by the function $\lambda/R$. Using this procedure, Prozorov found $R$ and analyzed the dependence of $R$ thus obtained on the aspect ratio of the cylinder. If similar calculations were carried out for the infinitely long slab, and the results were approximated by the function,
\begin{equation}\label{c2}
  \frac{\lambda}{R}+c_0\frac{\lambda^{2/3}}{R_{\rm crn}^{2/3}},
\end{equation}
where $R$ and $R_{\rm crn}$ are given by Eqs.~(\ref{16}) and (\ref{17}), this fit could give the value of $c_0$.

To understand the origin of the $\lambda^{2/3}$ term  in the magnetic moment, it is instructive to consider a slab which has a rounded edge. In other words, instead of one of the sharp corners of its cross sections,  we consider a quarter of the circle of the radius $\rho$ with $\lambda\ll \rho\ll w,d$. This relatively small change of the surface of the sample practically has no effect on the term $\lambda/R$ in Eq.~(\ref{15}), but a quarter of contribution $c_0(\lambda/R_{\rm crn})^{2/3}$ to the magnetic moment vanishes. However, instead of this quarter, a new term appears in the magnetic moment. This term $\delta M_y$ is caused by the renormalization of the surface sheet currents flowing near the arc of the length $\pi\rho/2$, and it can be estimated as follows:
\[
\delta M_y\sim \frac{H_a\lambda}{\rho}\left(\!\frac{(1-m)d}{6\sqrt{m}f(1,m)\rho} \right)^{\!1/3} \frac{\pi\rho}{2}w,
\]
where the $\partial H_t/\partial t$ in Eq.~(\ref{19}) is found with formula (\ref{b7}) for the cylinder of the radius $\rho$, and $H_t$ is calculated with expression (\ref{11}) at $l=\rho$. Using  expression (\ref{9}) for $M_y(\lambda=0)$, the magnetic moment per unit length of the slab with one rounded edge can be written in the form:
\begin{eqnarray}\label{c3}
 M_y(\lambda)=\! M_y(0)\!\left[1 -\frac{\lambda}{R}-\frac{c_1}{4}\frac{\lambda}{R_{\rm crn}^{2/3}\rho^{1/3}}-c_0\frac{3}{4}\frac{\lambda^{2/3}}{R_{\rm crn}^{2/3}} \right]\!,~~~~~
 \end{eqnarray}
where $R$ and $R_{\rm crn}$ are defined by Eqs.~(\ref{16}) and (\ref{17}), respectively, and $c_1\sim \pi/4$ according to the above rough estimate of $\delta M_y$. At $d/2w<1$ the linear-in-$\lambda$ contribution to $M_y$ is mainly determined by the term caused by the rounded edge of the slab. If $\rho$ decreases, this term increases, and at $\rho\sim \lambda$, it transforms into the nonanalytic term proportional to $\lambda^{2/3}$.

{}


\begin{thebibliography}{}


\bibitem{proz06} R. Prozorov and R.W. Giannetta, Magnetic penetration depth in unconventional superconductors, Supercond.\ Sci.\ Technol.\ {\bf 19}, R41 (2006).

\bibitem{hardy} W.N. Hardy, D. A. Bonn, D. C. Morgan, Ruixing Liang, and Kuan Zhang, Precision measurements of the temperature dependence of $\lambda$ in YBa$_2$Cu$_3$O$_{6.95}$: Strong evidence for nodes in the gap function, Phys.\ Rev.\ Lett. {\bf 70}, 3999 (1993).

\bibitem{carrington} A. Carrington, R.W. Giannetta, J. T. Kim, and J. Giapintzakis, Absence of nonlinear Meissner effect in YBa$_2$Cu$_3$O$_{6.95}$, Phys.\ Rev.\ B {\bf 59}, R14173 (1999).

\bibitem{proz11} R. Prozorov and V.G. Kogan, London penetration depth in iron-based superconductors, Rep.\ Prog.\ Phys.\ {\bf 74}, 124505 (2011).

\bibitem{putzke14} C. Putzke, P. Walmsley, J.D. Fletcher, L. Malone, D. Vignolles, C. Proust, S. Badoux, P. See, H.E. Beere,
D.A. Ritchie, S. Kasahara, Y. Mizukami, T. Shibauchi, Y. Matsuda \& A. Carrington, Anomalous critical fields in quantum critical superconductors, Nat.\ Commun.\ {\bf 5}, 5679 (2014).

\bibitem{shen} B. Shen, M. Leroux, Y.L. Wang, X. Luo, V.K. Vlasko-Vlasov, A.E. Koshelev, Z.L. Xiao, U. Welp, W.K. Kwok, M.P. Smylie, A. Snezhko, and V. Metlushko, Critical fields and vortex pinning in overdoped Ba$_{0.2}$K$_{0.8}$Fe$_2$As$_2$, Phys.\ Rev.\ B {\bf 91}, 174512 (2015).

\bibitem{juraszek20} J. Juraszek, R. Wawryk, Z. Henkie, M. Konczykowski, and T. Cichorek, Symmetry of order parameters in multiband superconductors LaRu$_4$As$_{12}$ and PrOs$_4$Sb$_{12}$ probed by local magnetization measurements, Phys.\ Rev.\ Lett. {\bf 124}, 027001 (2020).

\bibitem{yamashita} T. Yamashita, T. Takenaka, Y. Tokiwa, J.A. Wilcox,
Y. Mizukami, D. Terazawa, Y. Kasahara, S. Kittaka, T. Sakakibara,
M. Konczykowski, S. Seiro, H.S. Jeevan, C. Geibel, C. Putzke,
T. Onishi, H. Ikeda, A. Carrington, T. Shibauchi, Y. Matsuda, Fully gapped superconductivity with no sign change in the prototypical heavy-fermion CeCu$_2$Si$_2$,  Sci.\ Adv.\ {\bf 3}, e1601667 (2017).

\bibitem{takenaka} T. Takenaka, Y. Mizukami, J.A. Wilcox, M. Konczykowski, S. Seiro, C. Geibel, Y. Tokiwa, Y. Kasahara,
C. Putzke, Y. Matsuda, A. Carrington, and T. Shibauchi, Full-gap superconductivity robust against disorder in heavy-fermion CeCu$_2$Si$_2$, Phys.\ Rev.\ Lett.\ {\bf 119}, 077001 (2017).


\bibitem{pang} G. Pang, M. Smidman, J. Zhang, L. Jiao, Z. Weng, E.M. Nica, Y. Chen, W. Jiang, Y. Zhang, W. Xie, H.S. Jeevan, H. Lee, P. Gegenwart, F. Steglich, Q. Si, and H. Yuan, Fully gapped d-wave superconductivity in CeCu$_2$Si$_2$, PNAS {\bf 115}, 5343 (2018).


\bibitem{ishihara} K. Ishihara, M. Roppongi, M. Kobayashi, K. Imamura, Y. Mizukami, H. Sakai, P. Opletal, Y. Tokiwa,
Y. Haga, K. Hashimoto \& T. Shibauchi, Chiral superconductivity in UTe$_2$ probed by anisotropic low-energy excitations, Nat.\ Commun.\ {\bf 14}, 2966 (2023).

\bibitem{proz00} R. Prozorov, R. W. Giannetta, A. Carrington, F. M. Araujo-Moreira, Meissner-London state in superconductors of rectangular cross section in a perpendicular magnetic field, Phys.\ Rev.\ B {\bf 62}, 115 (2000).

\bibitem{giannetta} R. Giannetta, A. Carrington, R. Prozorov, J.\ Low Temp.\ Phys.\  {\bf 208}, 119-146 (2022).


\bibitem{klein} T. Klein, D. Braithwaite, A. Demuer, W. Knafo, G. Lapertot, C. Marcenat, P. Rodi\`ere, I. Sheikin, P. Strobel,
A. Sulpice, and P. Toulemonde, Thermodynamic phase diagram of Fe(Se$_{0.5}$Te$_{0.5}$) single crystals in fields up to 28 tesla, Phys.\ Rev.\ B {\bf 82}, 184506 (2010).


\bibitem{smylie} M. P. Smylie, H. Claus, U. Welp, W.-K. Kwok, Y. Qiu, Y. S. Hor, and A. Snezhko, Evidence of nodes in the order parameter of the superconducting doped topological insulator
Nb$_x$Bi$_2$Se$_3$ via penetration depth measurements Phys.\ Rev.\ B {\bf 94}, 180510(R) (2016).


\bibitem{prib} Z. Pribulov\'a, Z. Medveck\'a, J. Ka\v{c}mar\v{c}\'ik, V. Komanick\'y, T. Klein, P. Rodi\`ere, F. Levy-Bertrand, B. Michon, C. Marcenat, P. Husan\'ikov\'a, V. Cambel, J. \v{S}olt\'ys, G. Karapetrov, S. Borisenko, D. Evtushinsky, H. Berger, and P. Samuely, Magnetic and thermodynamic properties of Cu$_x$TiSe$_2$ single crystals,  Phys.\ Rev.\ B {\bf 95}, 174512 (2017).

\bibitem{shang18} T. Shang, G. M. Pang, C. Baines, W. B. Jiang, W. Xie, A. Wang, M. Medarde, E. Pomjakushina, M. Shi, J. Mesot, H. Q. Yuan, and T. Shiroka, Nodeless superconductivity and time-reversal symmetry breaking in the noncentrosymmetric superconductor Re$_{24}$Ti$_5$, Phys.\ Rev.\ B {\bf 97}, 020502(R) (2018).


\bibitem{kim1} H. Kim, K. Wang, Y. Nakajima, R. Hu, S. Ziemak,
P. Syers, L. Wang, H. Hodovanets, J.D. Denlinger,
P.M.R. Brydon, D.F. Agterberg, M.A. Tanatar,
R. Prozorov, J. Paglione,  Beyond triplet: Unconventional superconductivity in a spin-3/2 topological semimetal,
    Sci. Adv. {\bf 4}, eaao4513 (2018).

\bibitem{radmanesh} S.M.A. Radmanesh, C. Martin, Yanglin Zhu, X. Yin, H. Xiao, Z.Q. Mao, and L. Spinu,  Evidence for unconventional superconductivity in half-Heusler YPdBi and TbPdBi compounds revealed by London penetration depth measurements,
    Phys.\ Rev.\ B {\bf 98}, 241111(R) (2018).


\bibitem{kim} H. Kim, M.A. Tanatar, H. Hodovanets, K. Wang, J. Paglione, and R. Prozorov, Campbell penetration depth in low carrier density superconductor YPtBi, Phys.\ Rev.\ B {\bf 104}, 014510 (2021).

\bibitem{ishihara1} K. Ishihara, T. Takenaka, Y. Miao, Y. Mizukami, K. Hashimoto, M. Yamashita, M. Konczykowski, R. Masuki, M. Hirayama, T. Nomoto, R. Arita, O. Pavlosiuk, P. Wi\'sniewski, D. Kaczorowski, and T. Shibauchi, Tuning the parity mixing of singlet-septet pairing in a half-Heusler superconductor, Phys.\ Rev.\ X {\bf 11}, 041048 (2021).

\bibitem{wang21} A. Wang, Z.Y. Nie, F. Du, G.M. Pang, N. Kase, J. Akimitsu, Y. Chen, M.J. Gutmann, D.T. Adroja, R.S. Perry, C. Cao, M. Smidman, and H. Q. Yuan, Nodeless superconductivity in Lu$_{5-x}$Rh$_6$Sn$_{18+x}$ with broken time reversal symmetry, Phys.\ Rev.\ B {\bf 103}, 024503 (2021).


\bibitem{su} H. Su, Z.Y. Nie, F. Du, S.S. Luo, A. Wang, Y.J. Zhang, Y. Chen, P.K. Biswas, D. T. Adroja, C. Cao, M. Smidman, and H.Q. Yuan, Fully gapped superconductivity with preserved time-reversal symmetry in noncentrosymmetric LaPdIn, Phys.\ Rev.\ B {\bf 104}, 024505 (2021).

\bibitem{nie} Z.Y. Nie, J.W. Shu, A. Wang, H. Su, W.Y. Duan, A.D. Hillier, D.T. Adroja, P.K. Biswas, T. Takabatake, M. Smidman, and H.Q. Yuan, Nodeless superconductivity in noncentrosymmetric LaRhSn, Phys.\ Rev.\ B {\bf 105}, 134523 (2022).

\bibitem{duan} W. Duan, J. Zhang, R. Kumar, H. Su, Y. Zhou, Z. Nie, Y. Chen, M. Smidman, C. Cao, Y. Song, and H. Yuan, Nodeless superconductivity in the topological nodal-line semimetal CaSb$_2$, Phys.\ Rev.\ B {\bf 106}, 214521 (2022).

\bibitem{chapai} R. Chapai, M.P. Smylie, H. Hebbeker, D.Y. Chung, W.-K. Kwok, J.F. Mitchell, and U. Welp, Superconducting properties and gap structure of the topological superconductor candidate Ti$_3$Sb, Phys.\ Rev.\ B {\bf 107}, 104504 (2023).


\bibitem{proz_pr-appl21} R.\ Prozorov, Meissner-London susceptibility of superconducting right circular cylinders in an axial magnetic field, Phys.\ Rev.\ Appl.\ {\bf 16}, 024014 (2021).

\bibitem{Meissner} E.H. Brandt, G.P. Mikitik, Meissner-London currents in superconductors with rectangular cross section, Phys.\ Rev.\ Lett. {\bf 85}, 4164 (2000).

\bibitem{LL} L.D. Landau and E.M. Lifshitz, {\it Electrodynamics
of Continuous Media}, Course in Theoretical Physics Vol.~8
(Pergamon, Oxforf-London-New York-Paris, 1960).

\bibitem{prb21} G.P. Mikitik, Critical current in thin flat superconductors with Bean-Livingston and geometrical barriers, Phys.\ Rev.\ B {\bf 104}, 094526 (2021).


\bibitem{jetp13} E.H. Brandt, G.P. Mikitik, E. Zeldov, Two regimes of vortex penetration into platelet-shaped type-II superconductors, JETP {\bf 117}, 439 (2013).

\bibitem{m-sh24} G.P. Mikitik and Yu.V. Sharlai, Determination of the magnetic penetration depth with measurements of the vortex-penetration field for type-II superconductors, Phys.\ Rev.\ B {\bf 110}, 064507  (2024).

\bibitem{br94} E.H. Brandt, Thin superconductors in a perpendicular magnetic ac field: General formulation and strip geometry,
Phys.\ Rev.\ B {\bf 49}, 9024 (1994).

\bibitem{br04} G.P. Mikitik, E.H. Brandt, and M. Indenbom, Superconducting strip in an oblique magnetic field, Phys.\ Rev.\ B {\bf 70}, 014520 (2004).

\bibitem{bl} G. Blatter, M.V. Feigel'man, V.B. Geshkenbein,
 A.I. Larkin, and V.M. Vinokur, Vortices in high-temperature superconductors,  Rev.\ Mod.\ Phys.\ {\bf 66}, 1125 (1994).


\bibitem{proz-apl} R.\ Prozorov, R.W.\ Giannetta, A.\ Carrington, P.\ Fournier, R.L.\ Greene, P.\ Guptasarma, D.G.\ Hinks, and A.R.\
    Banks, Measurements of the absolute value of the penetration depth in high-$T_c$ superconductors using a low-$T_c$ superconductive coating, Appl.\ Phys.\ Lett.\ {\bf 77}, 4202 (2000).


\bibitem{osborn} J.A. Osborn, Demagnetizing factors of the general ellipsoid, Phys.\ Rev.\ {\bf 67}, 351 (1945).

\end{thebibliography}
\end{document}